\documentclass[floatfix,superscriptaddress,amsmath,amssymb,longbibliography,twocolumn]{revtex4-2}
\usepackage{graphicx}
\usepackage{color}
\usepackage{hyperref}
\hypersetup{colorlinks=true,citecolor=blue,linkcolor=blue,urlcolor=blue}
\allowdisplaybreaks

\begin{document}

\title{
Observation of anomalous tunneling in collective excitations via a cloud experiment platform for Bose-Einstein condensates
}

\author{Daichi Kagamihara}
\email{dkagamihara119@g.chuo-u.ac.jp}
\affiliation{Department of Physics, Chuo University, Bunkyo, Tokyo 112-8551, Japan}

\author{Hironori Kazuta}
\affiliation{Department of Physics, Kindai University, Higashi-Osaka, Osaka 577-8502, Japan}

\author{Yewei Wu}
\affiliation{Infleqtion, 3030 Sterling Cir., Boulder, CO 80301, USA}

\author{N. J. Fitch}
\affiliation{Infleqtion, 3030 Sterling Cir., Boulder, CO 80301, USA}

\author{Ippei Danshita}
\email{danshita@phys.kindai.ac.jp}
\affiliation{Department of Physics, Kindai University, Higashi-Osaka, Osaka 577-8502, Japan}

\date{\today}

\begin{abstract}
Recent development of cloud-based experiment platforms has enabled physicists to examine theoretical concepts with unprecedented accessibility. Oqtant is a cloud-accessible platform for trapped Bose-Einstein Condensates (BECs) of neutral atomic gases, providing an invaluable experimental tool for studying the dynamics of BECs. 
An intriguing theoretical prediction of a characteristic phenomenon of BECs is anomalous tunneling, whereby low-energy phonon excitations of BECs easily transmit through a barrier potential. We utilize Oqtant to observe the effects of anomalous tunneling on collective excitations of BECs. For this purpose, we theoretically show that anomalous tunneling affects the frequencies of the collective excitations in the low-energy regime, and experimentally measure these frequencies using Oqtant. Our results reveal that low-energy collective modes are less affected by a potential barrier, which indicates the presence of anomalous tunneling. Our work would contribute to fundamental understandings of BECs, as well as highlight the potential of cloud-based experiments in quantum-body physics.
\end{abstract}

\maketitle


\section{Introduction.}
\label{sec_introduction}

In recent years, the emergence of cloud-based experiment platforms has significantly expanded research methodologies in physics. This trend has been particularly prevalent in the area of quantum computing, with IBM Quantum\,\cite{IBMQ}, Amazon Braket\,\cite{Amazonbraket}, and Google Quantum AI\,\cite{google} serving as prominent examples. These services allow researchers to conduct experiments remotely without the need for direct access to expensive hardware. These cloud platforms facilitate the exploration of new phenomena and the rapid testing of theoretical concepts.

Oqtant is one such service, a cloud-accessible platform for a Bose-Einstein condensate (BEC) of ultracold neutral atomic gas, which is provided by a company in the USA, namely Infleqtion\, \cite{Oqtantweb}.  In the Oqtant platform, one can create a BEC of $^{87}{\rm Rb}$ atoms trapped in a cigar-shaped potential and utilize many useful functionalities. In particular, one can manipulate in real time repulsive potentials for atoms generated by blue-detuned lasers and observe the time evolution of the density distribution of atoms. This capability allows even theoretical researchers to experimentally explore the dynamics of a weakly-interacting dilute Bose gas. 

As an interesting phenomenon in a weakly interacting BEC, anomalous tunneling of Bogoliubov excitations, which are elementary excitations of the BEC, was theoretically predicted in the 2000s\,\cite{Kovrizhin:2001aa,Kagan:2003aa}. When one considers a one-dimensional (1D) scattering problem of Bogoliubov excitations across a repulsive potential barrier illustrated in Fig.~\ref{fig_trans_amp}(a), the tunneling probability is higher for lower excitation energies. At the zero-energy limit, it even exhibits perfect transmission. This behavior is in stark contrast with the tunneling of a single particle, so it is called anomalous tunneling. This phenomenon is related to superfluidity\,\cite{Danshita:2006aa,Danshita:2007aa,Takahashi:2009aa} and is conjectured to be a universal property of Nambu-Goldstone modes associated with the spontaneous breaking of continuous symmetry\,\cite{Watabe:2011aa,Kato:2012aa,Nakayama:2015aa}. Despite such extensive theoretical interest, experimental observation of anomalous tunneling has not been realized so far.

In this paper, using the Oqtant platform, we experimentally observe some signatures of anomalous tunneling in collective excitation modes of BECs in double-well potentials consisting of a trap potential and a repulsive potential barrier at the trap center. Previous theoretical work has shown that anomalous tunneling is closely related to how the lowest frequency of collective modes of BECs in the double-well potentials depends on the barrier height~\cite{Danshita:2005aa}. To provide theoretical references more specialized to the experimental setup of Oqtant, we theoretically extend this analysis to several higher-frequency modes. We show that anomalous tunneling is also relevant to the barrier-height dependence of the higher mode frequencies. In particular, in addition to the lowest frequency of the collective mode, the second and third lowest frequencies are suitable for measurement within the specification of Oqtant. We also report experimental results obtained using Oqtant. In our experiment setup, we apply the repulsive potential to excite collective modes and then observe the time evolution of the density distribution of atom clouds after removing the potential to evaluate the frequencies of collective modes. The above protocol is performed in the presence of a potential barrier for various barrier heights. We discuss the relation between the observed barrier-height dependence of collective mode frequencies and the anomalous tunneling effect.

\section{Results}
\label{sec_theory}

\subsection{Theoretical analysis}

In order to conduct numerical simulations with physical parameters of BECs of the Oqtant platform, let us first introduce the specifications of Oqtant. Rubidium $^{87}\mathrm{Rb}$ atoms in the hyperfine state $|F=2,m_F=2\rangle$ are trapped by the cigar-shaped harmonic potential $V_{\mathrm{trap}} = m [\omega_{\perp}^2 (x^2 + y^2) + \omega_z^2 z^2] / 2$, where $m$ is the mass of Rubidium atoms. The transverse and axial trap frequencies are, respectively, given by $\omega_{\perp} = 2\pi \times 400 \mathrm{Hz}$ and $\omega_{z} = 2\pi \times 42 \mathrm{Hz}$. The value of $\omega_{\perp}$ is taken from the Oqtant manual \cite{Oqtantweb}, whereas that of $\omega_z$ is determined from the dipole motion experiment. BECs are produced via evaporative cooling. The typical temperature of BECs is about 90nK, and the condensate atom number $N_{\rm c}$ is about 8300 (for more details, see Supplementary Note 1). We can measure the atom density distribution until 100ms after BEC preparation.  

Since $\omega_{\perp} \gg \omega_{z}$, the condensate is expected to show quasi-1D behavior. Although actual experiments are at finite temperatures, we ignore finite temperature effects in our theoretical analysis. Thus, we use the effective quasi-1D Gross-Pitaevskii (GP) equation~\cite{Gerbier2004, Mateo:2007aa, Middelkamp:2010aa};
\begin{align}
i \hbar \frac{\partial \psi(z,t)}{\partial t} = \left(- \frac{\hbar^2}{2m}\frac{\partial^2}{\partial z^2} + V_{\mathrm{ext}}(z) + \hbar \omega_{\perp} \sqrt{1 + 4 a |\psi(z,t)|^2} \right)\psi(z,t),
\label{eq_q1DGPeq}
\end{align}
where $\psi(z,t)$ is the 1D condensate wave function, which is normalized as $\int dz |\psi(z,t)|^2 = N_{\rm c}$. $V_{\mathrm{ext}}(z)$ represents the external potential term along the $z$ direction, including the trap potential and the other potentials that we input.

To analyze anomalous tunneling of excitations through a barrier and collective excitations in a double-well potential, we assume that the condensate is weakly perturbed. We put the solution of Eq.\,\eqref{eq_q1DGPeq} as $\psi(z,t) = \left( \psi_0(z) + u(z) e^{-i\varepsilon t / \hbar} - v^*(z) e^{i\varepsilon t / \hbar} \right) e^{-i\mu t / \hbar}$, where $\psi_0(z)$ is the static solution of Eq.\,\eqref{eq_q1DGPeq}, $(u(z), v(z))$ the small fluctuations, $\mu$ the chemical potential, and $\varepsilon$ the excitation energy. The static solution $\psi_0(z)$ satisfies the time-independent GP equation 
\begin{align}
\left( \mathcal{L} + \hbar \omega_{\perp} \sqrt{1 + 4 a |\psi_0(z)|^2} \right) \psi_0(z) = 0,
\label{eq_time_indep_GP}
\end{align}
where $\mathcal{L} = -\frac{\hbar^2}{2m}\frac{\partial^2}{\partial z^2} - \mu + V_{\mathrm{ext}}(z)$, and the fluctuations $u(z)$ and $v(z)$ are determined by the linearized GP equation (Bogoliubov equation)\,\cite{Middelkamp:2010aa}
\begin{align} \nonumber
\left[ \mathcal{L} + f(z) \right] u - gv =& \varepsilon u,
\\
\left[ \mathcal{L}  + f(z) \right] v - gu =& - \varepsilon v,
\label{eq_Bogoliubov_eq}
\end{align}
where $f(z) = |g(z)| + \hbar \omega_{\perp} \sqrt{1+4a |\psi_0(z)|^2}$ and $g(z) = 2\hbar \omega_{\perp} a\psi_0^2(z) / \sqrt{1+4a|\psi_0(z)|^2}$.

\begin{figure}[t]
\centering
\includegraphics[width=\columnwidth]{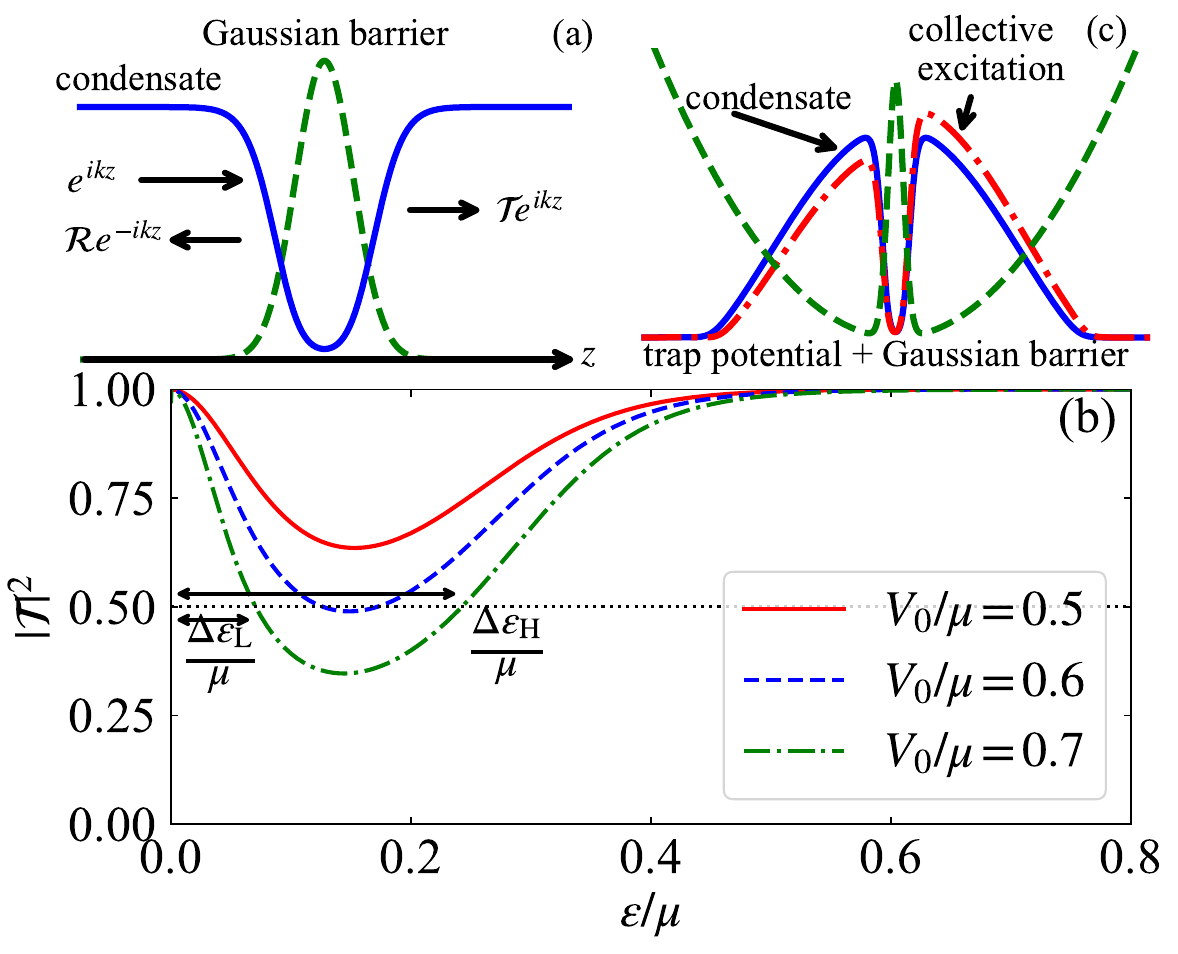}
\caption{(a) Schematic picture of the tunneling problem. The solid and dashed curves mean the condensate density $|\psi_0|^2$ and the Gaussian barrier. (b) Calculated transmission coefficient $|\mathcal{T}|^2$ as functions of the energy of the incident wave $\varepsilon$ for some barrier heights $V_0$. We also show that the halfwidths $\Delta \varepsilon_{\rm L}$ and $\Delta \varepsilon_{\rm H}$ for $V_0 / \mu = 0.7$. (c) Schematic drawing of the situation in examining collective modes. The solid and dashed curves are the same as panel (a). The dash-dot line denotes a collective excitation.
}
\label{fig_trans_amp}
\end{figure}

To check how anomalous tunneling occurs in the current quasi-1D situation, we consider a tunneling problem as schematically shown in Fig.\,\ref{fig_trans_amp}(a). We set $V_{\rm ext}$ a Gaussian potential; $V_{\mathrm{ext}} = V_0 \exp\left(-z^2 / 2\sigma^2\right)$. The width $\sigma$ is set to be 0.5$\mu$m, which is the smallest one available in Oqtant. 

We show the energy dependence of the transmission coefficient $|\mathcal{T}|^2$ in Fig.\,\ref{fig_trans_amp}(b). As the energy is lowered from the high-energy side, the transmission coefficient first decreases but then increases in the low-energy region, finally approaching unity at the zero-energy limit. This is nothing but the anomalous tunneling effect.

\begin{figure}[t]
\centering
\includegraphics[width=\columnwidth]{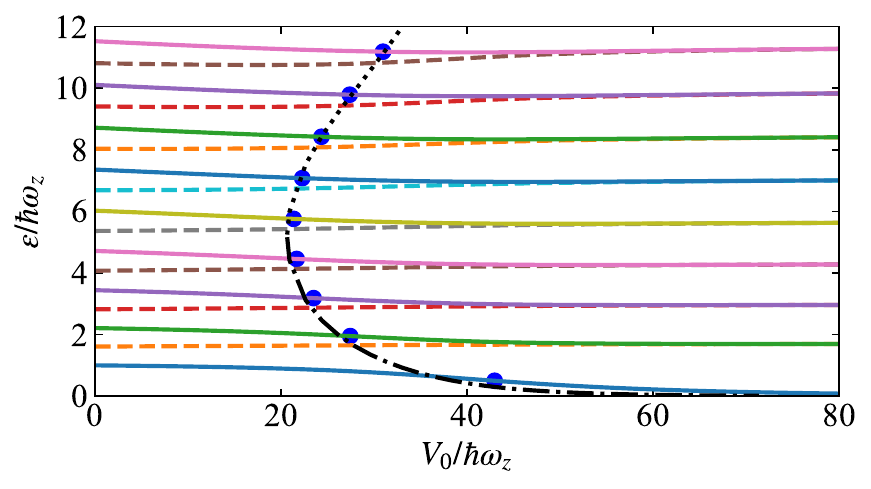}
\caption{Calculated collective mode frequencies $\varepsilon$ as functions of the barrier height $V_0$. The solid and dashed lines are odd-parity mode and even-parity mode frequencies. The dash-dotted line shows the halfwidth due to anomalous tunneling $\Delta \varepsilon_{\rm L}(V_0)$. The dotted line is the halfwidth of high-energy excitations $\Delta \varepsilon_{\rm H}(V_0)$. We also put dots at $(V_{\rm m}, \varepsilon_{2i+1}(V_{\rm m}))$ to indicate characteristic barrier height $V_{\rm m}$ for the merging of two neighboring modes.
}
\label{fig_freqs_de}
\end{figure}

A higher barrier potential $V_0$ induces a more abrupt increase in the transmission coefficient, as seen in Fig.\,\ref{fig_trans_amp}(b). For convenience in discussing the relation between anomalous tunneling and collective excitation frequencies, we define the halfwidth $\Delta \varepsilon$ of the transmission coefficient, i.e., the energy at which the transmission coefficient equals 0.5. We note that there exist two halfwidths, $\Delta \varepsilon_{\mathrm{L}}$ and $\Delta \varepsilon_{\mathrm{H}}$, as seen in Fig.\,\ref{fig_trans_amp}(b). The low-energy halfwidth $\Delta \varepsilon_{\mathrm{L}}$ is due to anomalous tunneling, but the high-energy one $\Delta \varepsilon_{\mathrm{H}}$ simply comes from the well-known fact that high-energy excitations are less sensitive to a potential. These halfwidths are plotted as functions of $V_0$ in Fig.\,\ref{fig_freqs_de}. Excitations with energy $\varepsilon < \Delta \varepsilon_{\mathrm{L}}$ or $\varepsilon > \Delta \varepsilon_{\mathrm{H}}$ are not significantly affected by a barrier.

Next, we investigate the impact of anomalous tunneling on collective excitations under the Oqtant setup. We set the potential as $V_{\mathrm{ext}} = m\omega_z^2 z^2/2 + V_0 \exp\left(-z^2 / 2\sigma^2\right)$ and solve Eqs.\,\eqref{eq_time_indep_GP} and \eqref{eq_Bogoliubov_eq} to obtain the eigen frequencies $\varepsilon/\hbar$ of collective excitations. The situation is schematically depicted in Fig.\,\ref{fig_trans_amp}(c).

We show calculated collective mode frequencies as functions of the barrier height in Fig.\,\ref{fig_freqs_de}. We see that the two neighboring frequencies $\varepsilon_{2i}(V_0)$ and $\varepsilon_{2i+1}(V_0)$ $(i = 0,1,2,\dots)$ tend to merge as $V_0$ is increased. Notice that the lowest frequency $\varepsilon_{1}$ tends to merge with the zero mode one $\varepsilon_0 = 0$. This tendency is easily understood: When the barrier height is high enough, the condensate can be regarded as two independent BECs. In this case, the in-phase and out-of-phase oscillations of two BECs should have the same frequency. The even collective mode $\varepsilon_{2i}$ (odd one $\varepsilon_{2i+1}$) is indeed an even (odd) parity mode and then converges to out-of-phase (in-phase) oscillation at $V_0 \gg \mu$.

We introduce a characteristic barrier height $V_{\mathrm{m}}$ for the merging of two collective modes. We define $V_{\mathrm{m}}$ at which the difference of frequencies between two neighboring modes is half of that in the absence of the barrier; $V_{\mathrm{m}}$ satisfies $[\varepsilon_{2i+1}(V_{\mathrm{m}}) - \varepsilon_{2i}(V_{\mathrm{m}})] / [\varepsilon_{2i+1}(0) - \varepsilon_{2i}(0)] = 1/2$, which is shown by the dots in Fig.\,\ref{fig_freqs_de}. We also show the halfwidth $\Delta \varepsilon(V_0)$ in Fig.\,\ref{fig_freqs_de} and find that $V_{\mathrm{m}}$ and $\Delta \varepsilon(V_0)$ are well close to each other.

The coincidence between $\Delta \varepsilon(V_0)$ and $V_{\mathrm{m}}$ suggests an interpretation of the merging behavior. In the region left of $\Delta \varepsilon(V_0)$ in Fig.\,\ref{fig_freqs_de}, the transmission coefficient is greater than $1/2$, so the effects of the barrier on the collective excitations are expected to be insignificant. Conversely, in the region right of $\Delta \varepsilon(V_0)$, the barrier strongly affects the collective excitations, making two collective modes merge in the high barrier height limit. In this way, the merging behavior is also characterized by $\Delta \varepsilon(V_0)$, which explains why $V_{\rm m}$ is close to $\Delta \varepsilon(V_0)$.

Now, we discuss how to capture anomalous tunneling from measurements of collective mode frequencies. The consequence of anomalous tunneling is that $\Delta \varepsilon(V_0)$ tends to correspond to a large $V_0$ in the low-energy region. Since $\Delta \varepsilon(V_0)$ and $V_{\mathrm{m}}$ are close to each other, the observation of the merging barrier height $V_{\rm m}$ for $i=0$ and $i=1$ and of the relation $V_{{\rm m},i=0} > V_{{\rm m},i=1}$ would be indirect evidence of anomalous tunneling. The typical time scales of the first and second (third) lowest collective excitations are, respectively, $T_{1} = 2\pi / \omega_z \approx 24\mathrm{ms}$ and $T_{2(3)} \approx 15\mathrm{ms}$, which are feasible in Oqtant.

\subsection{Experimental results}

\begin{figure}[t]
\centering
\includegraphics[width=\columnwidth]{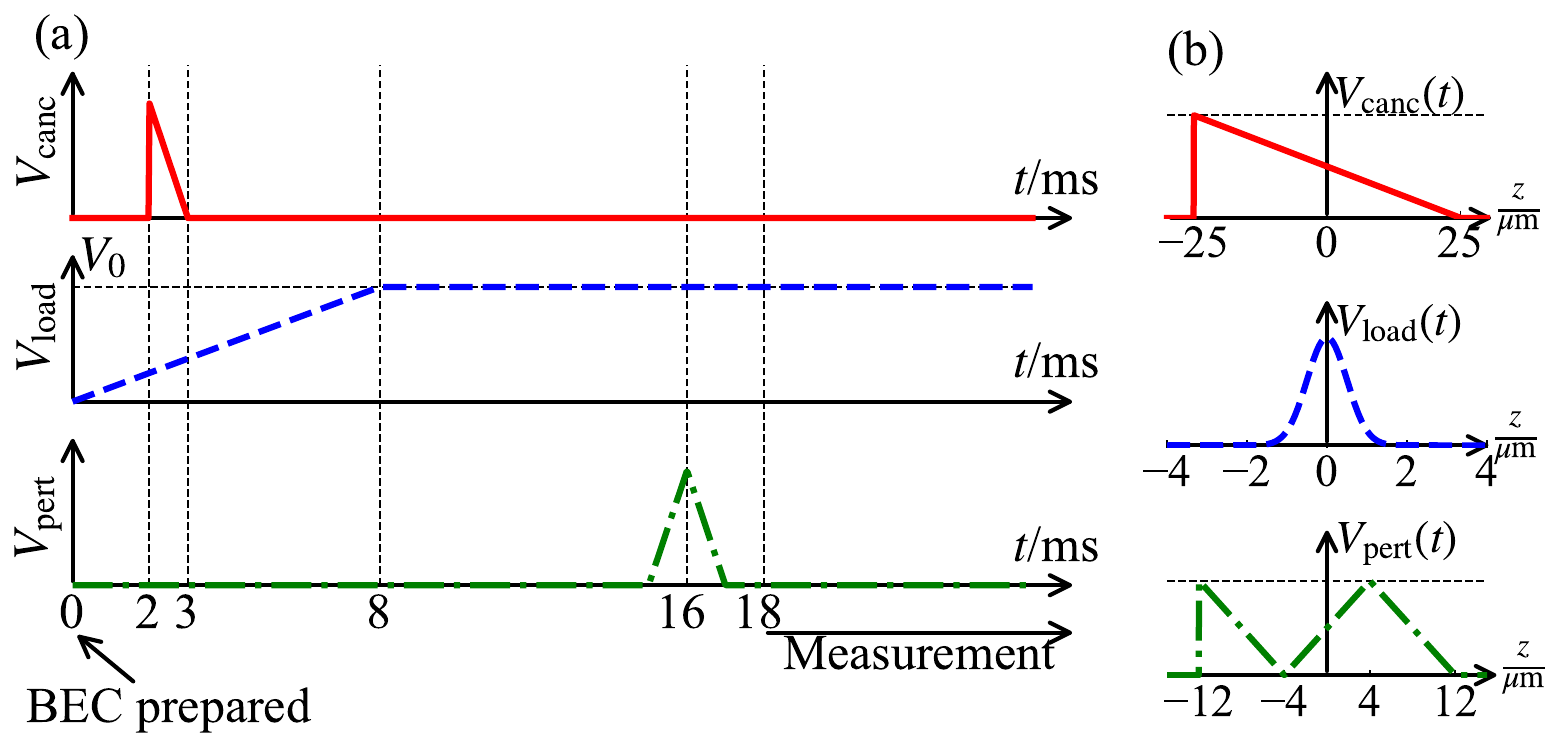}
\caption{Experimental protocol after BEC preparation. (a) Barrier schedule. We use three potentials: $V_{\rm canc}$, $V_{\rm load}$, and $V_{\rm pert}$. The horizontal axis represents elapsed time after BEC preparation. The vertical axis shows potential heights. (b) Potential shape. $V_{\rm canc}$ is used for the cancellation of unwanted initial motion. $V_{\rm load}$ is the barrier at the center of the trap and linearly grows to $V_0$ over 8ms. $V_{\rm pert}$ is used to excite collective motions. We actually use three types of perturbation potentials and here show only one of them; for more details, see Supplementary Note 2.
}
\label{fig_experiment_protocol}
\end{figure}

To observe low-energy collective excitations by using Oqtant, we perform experiments shown in Fig.\,\ref{fig_experiment_protocol} after preparing a BEC in the cigar-shaped trap mentioned above. The prepared BEC cloud is in the dipole motion. To reduce this unwanted oscillation, we add a linear potential from 2 ms to 3 ms (0 ms corresponds to the end of the BEC preparation). We ramp up the barrier at the center of the trap linearly in time up to $V_0$ over 8 ms to create a double-well potential. Notice that a BEC in a double-well potential has also been created in previous experiments~\cite{Andrews:1997aa,Albiez:2005aa,Schumm:2005aa,Jo:2007aa,LeBlanc:2011aa}. While a few low-energy collective excitations have been observed in Ref.~\cite{Albiez:2005aa,LeBlanc:2011aa}, their connection to anomalous tunneling has never been addressed. From 15 ms to 17 ms, we add a perturbation potential to excite collective motions. We use three types of potentials depending on the height of the barrier and which collective mode we intend to excite (for more details, see Supplementary Note 2). After removing the perturbation potential, we measure the atom cloud distribution in time.

\begin{figure}[t]
\centering
\includegraphics[width=\columnwidth]{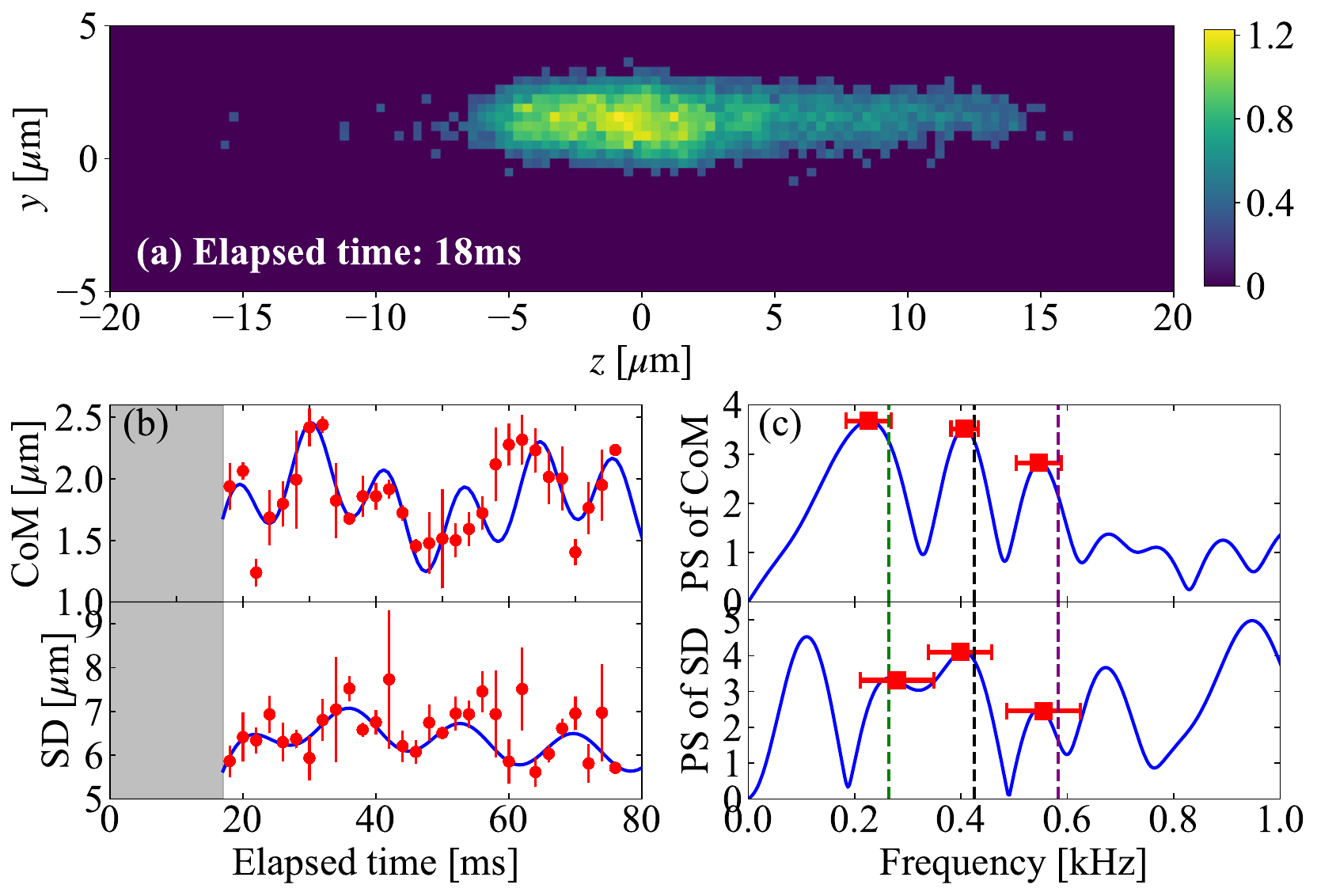}
\caption{(a) One example of observed atom cloud distribution at elapsed time 18 ms in the absence of a barrier. The color bar shows the optical depth corresponding to the atom density in an arbitrary unit. (b) Center of mass (CoM) and standard deviation (SD) of atom clouds along the $z$ direction. We also show these power spectra (PS) in panel (c). In the power spectra, we remove the offsets of the original signals. The squares with error bars show the estimated peaks of PS. The vertical dashed lines show the theoretical predictions for the lowest, second-lowest, and third-lowest collective mode frequencies in descending order of frequency.
}
\label{fig_experiment_result}
\end{figure}

Figure \ref{fig_experiment_result}(a) shows an example of observed atom cloud distributions after removing background noise. To extract the frequency of collective excitations, we calculate the center of mass position (CoM) and the standard deviation (SD), as shown in Fig.\,\ref{fig_experiment_result}(b). We mainly use CoM (SD) for the first and third (second) mode frequency estimation because it is expected to be sensitive to odd (even) parity modes.

We determine frequencies and their error in two ways. The first method is fitting of the observed time evolution of CoM and SD with the function $f(t) = A_1 e^{-B_1 t} \sin(C_1 t + D_1) + A_2 e^{-B_2 t} \sin(C_2 t + D_2) + E$. We take the $C_{1(2)}$ exhibiting larger oscillation amplitudes $A_{i}$ as estimates of the lowest mode. The $C_1$ or $C_2$ value close to the theoretical prediction is taken as an estimated frequency of the second and third modes. The error is estimated from the standard deviation of $C_{1(2)}$.

The second method is a peak analysis in the Fourier spectra. Figure \ref{fig_experiment_result}(c) shows the Fourier spectra of CoM and SD (blue curves), and the peaks (red squares with error bars). We also plot the theoretical predictions for the frequencies of the lowest three collective excitations as the vertical lines in Fig.\,\ref{fig_experiment_result}(c). We use the frequency and standard deviation of the dominant peak as estimates of the lowest mode. In addition, we use those of the peak whose frequency is close to the theoretical prediction as estimates of the second and third modes.

\begin{figure}[t]
\centering
\includegraphics[width=\columnwidth]{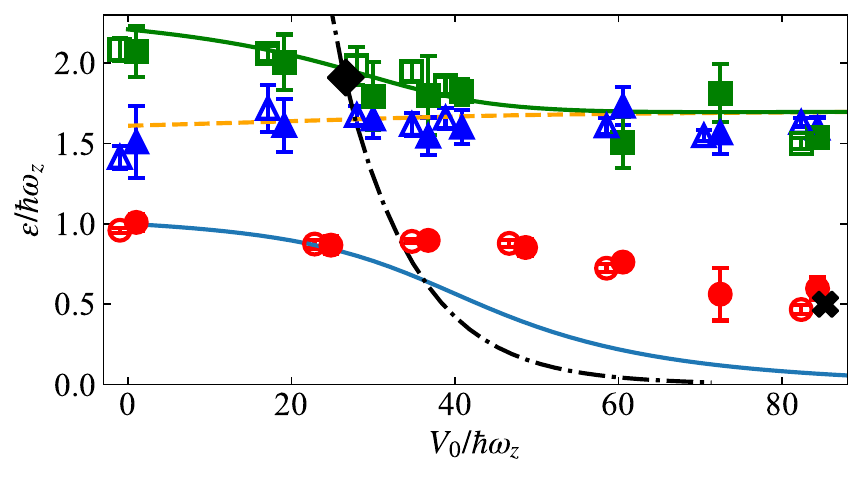}
\caption{Estimated collective mode frequency. Circles, triangles, and squares denote the estimated lowest, second-lowest, and third-lowest frequencies, respectively. Unfilled (filled) symbols represent frequencies estimated from fit (Fourier) analysis, whose $x$-axis values are shifted by $-1(+1)$ for better readability.  The data of the lowest mode from fitting at $V_0 / \hbar \omega_z \approx 70$ and those of the third-lowest mode from fitting at $V_0 / \hbar \omega_z \approx 60$ and $70$ are absent because we cannot determine them. Solid, dashed, and dash-dotted lines are the same as those in Fig.\,\ref{fig_freqs_de}. The cross and diamond show the collective mode frequency merging points for the lowest and second-third lowest modes determined from experimental data, respectively.
}
\label{fig_experiment_freq_summary}
\end{figure}

The estimated frequencies are summarized in Fig.\,\ref{fig_experiment_freq_summary}. As for the second and third collective excitations, the measured frequencies agree well with the theoretical predictions, i.e., with increasing the barrier height, the two frequencies approach each other. We also find that the merging behavior becomes significant when the excitation energy $\varepsilon$ is larger than the halfwidth of anomalous tunneling $\Delta \varepsilon$.

As for the lowest-frequency collective excitations, experimental results and theoretical predictions show the same qualitative tendency: As the barrier height increases, both decrease. Although they quantitatively show a large deviation compared to the second and third collective excitation cases, we obtain that the lowest mode merging barrier height ($V_{{\rm m},i=0}^{\rm exp} \sim3.58$kHz) is larger than the second-third mode one ($V_{{\rm m},i=1}^{\rm exp} \sim 1.01$kHz), which is also a signature of anomalous tunneling.

Let us briefly discuss the discrepancy between theory and experiment of the lowest-frequency mode. Finite-temperature effects may be responsible for this. However, finite-temperature analysis in a cigar-shaped and double-well trap\,\cite{Xhani:2022aa} shows that finite-temperature effects reduce the frequency of the lowest-frequency mode. The tendency is the opposite since our measured frequency data are larger than the zero-temperature theoretical data. To solve this problem, more sophisticated frequency analysis, such as machine learning\,\cite{Wu:2020aa}, and careful setup suitable for lowest-frequency mode detection would be helpful, but we leave these as future problems.

\section{Discussion}
\label{sec_summary}

We theoretically and experimentally studied the collective excitation modes of a BEC in a double-well potential using the modified GP equation for quasi-1D geometry and the cloud-accessible BEC platform named Oqtant. In the theory part, we showed that two neighboring collective mode-frequencies merge as the barrier height increases, and that the merging is related to anomalous tunneling in the low-energy region. In the experiment part, we observed the merging behavior of collective excitations by using Oqtant. The observed frequencies of the second and third collective modes agree well with theoretical predictions. As for the lowest collective modes, the measured frequency is considerably deviated from the theoretical prediction, but has the same qualitative tendency: As the barrier increases, their frequencies decrease. Moreover, we confirmed that experimentally determined merging points satisfy $V^{\rm exp}_{{\rm m},i=1} < V^{\rm exp}_{{\rm m},i=0}$. From these observations, we conclude that we captured indirect evidence of anomalous tunneling in our Oqtant experiment.

As a future direction, it would be exciting to perform other interesting BEC experiments using Oqtant. In doing so, since the Oqtant setup is at finite temperatures, the simulation by the Zaremba-Nikuni-Griffin equation\,\cite{Griffin:2009aa} would be useful. Various cloud experiment platforms have recently become available, such as Aquila (Rydberg atom system by QuEra)~\cite{Wurtz:2023aa}. The use of these platforms is another interesting future direction (the use of Aquila has already been reported\,\cite{Dag:2024aa,Bauer:2024aa,Kaufman:2025aa}).

\section{Methods}

\subsection{Theoretical analysis of tunneling coefficient}

In a tunneling problem as schematically shown in Fig.\,\ref{fig_trans_amp}(a), we should specify the chemical potential $\mu$ rather than $N_{\rm c}$, because we are considering the infinite system. We use the chemical potential as $\mu / (\hbar\omega_z) \approx 34.12$, which is determined by solving Eq.\,\eqref{eq_time_indep_GP} with the trap potential $V_{\rm ext} = m\omega_z^2z^2/2$ and $N_{\rm c} = 8300$.

The static condensate $\psi_0$ is determined from Eq.\,\eqref{eq_time_indep_GP}. The plane waves of fluctuations $u$ and $v$ come from $z \to -\infty$ and then are scattered by the Gaussian potential. The situation can be taken into account by setting the boundary condition as
\begin{align}
\begin{pmatrix} u(z) \\ v(z) \end{pmatrix} = \begin{cases} \begin{pmatrix} \tilde{u} \\ \tilde{v} \end{pmatrix} e^{ikz} + \mathcal{R} \begin{pmatrix} \tilde{u} \\ \tilde{v} \end{pmatrix} e^{-ikz} & z \to -\infty
\\
\mathcal{T} \begin{pmatrix} \tilde{u} \\ \tilde{v} \end{pmatrix} e^{ikz} & z \to \infty
\end{cases},
\label{eq_boundary_cond}
\end{align}
where the wave number $k$ of the incident wave satisfies the dispersion relation of the Bogoliubov equation in the uniform regime; $\varepsilon^2 = \frac{\hbar^2 k^2}{2m} \left( \frac{\hbar^2 k^2}{2m} + \frac{4\hbar \omega_{\perp} a n_0}{\sqrt{1+4an_0}} \right)$ with the uniform condensate density $n_0= \lim_{|z|\to\infty} |\psi_0(z)|^2$. $\mathcal{T}$ and $\mathcal{R}$ in Eq.\,\eqref{eq_boundary_cond} represent the transmission and reflection amplitudes satisfying $|\mathcal{T}|^2 + |\mathcal{R}|^2=1$.

\subsection{Peak analysis of Fourier spectra of experimental data}

In analyzing peaks of Fourier spectra of the center of mass (CoM) and standard deviation (SD) of observed atom cloud distributions, statistical errors should be treated carefully.
To account for the influence of statistical errors on peaks, we calculate the averages and standard deviations of CoM and SD for each time and randomly sample them according to the Gaussian distribution. We generate 1024 sampled time series data and calculate their Fourier spectra. We also applied zero-padding to the input signal before performing the Fourier transform to improve the frequency resolution. For each peak in the Fourier spectra, we evaluate sample averages of peak frequencies and their standard deviation. Figure 4(c) in the main text shows the sample-averaged Fourier spectra of CoM and SD (blue curves), and the sample-averaged peaks (red squares with error bars).

\subsection{Extract merging points from experimental data}

To determine merging points from experimental data, we apply linear fitting to data near merging points and evaluate $V_{\rm m}$ from it. For the lowest-frequency mode, we use $V_0 / \text{kHz}$ = 2.5, 3.0, 3.5 data. For the second-third-lowest frequency merging point, we use $V_0 / \text{kHz}$ = 0.7605, 1.207, 1.5 data. 

\section*{Data availability}

The experimental and simulation datasets that support the findings of this study are openly available in Zenodo at DOI:\doi{10.5281/zenodo.17336579}.

\section*{Code availability}

The scripts used for data analysis and simulation are also included in the Zenodo repository (DOI:\doi{10.5281/zenodo.17336579}).

%

\section*{Acknowledgments}

The authors acknowledge collaborations in the early stage of this work with Takayuki Tanaka and Hibiki Takegami. They also thank Takafumi Tomita, Shintaro Taie, and Shin Inouye for useful discussions. This work was financially supported by JSPS KAKENHI (Grant No.\ JP25K17319), MEXT Q-LEAP (Grant No.\ JPMXS0118069021), JST FOREST (Grant No.\ JPMJFR202T), and JST ASPIRE (Grant No.\ JPMJAP24C2).

\section*{Author contributions.}

D.K.\ and I.D.\ performed theoretical calculations, designed experimental protocols, and analyzed experimental data. D.K., H.K., and I.D.\  remotely conducted the experiments through the cloud. Y.W.\ and N.F.\ developed the hardware of Oqtant. D.K. and I.D. contributed to the writing of the paper.

\section*{Competing interests.}

Y.W.\ and N.F.\ are employed by Infleqtion, whose cloud service was used for the experiment in this work. The remaining authors declare no other competing interests.

\clearpage
\onecolumngrid

\renewcommand{\thefigure}{S\arabic{figure}}
\renewcommand{\thetable}{S\arabic{table}}
\setcounter{figure}{0}
\setcounter{table}{0}

\appendix
\begin{center}
{
\large
Supplemental Information: ``Observation of anomalous tunneling in collective excitations via a cloud
experiment platform for Bose-Einstein condensates''
}
\end{center}

In this supplemental information, we explain (i) parameters of prepared BECs and (ii) details of collective mode measurement.

\subsection*{Supplementary Note 1: Evaporation process and BEC parameters}

The experiment was conducted every two weeks on Tuesdays, Wednesdays, and Thursdays from April 2024 to October 2024. In July 2024, Oqtant was recalibrated with a change of barrier calibration. After that, the temperature of BECs changed even with the same evaporation cooling process parameters as before. For this reason, we tuned the evaporation cooling process after the recalibration to make the difference in parameters of BECs between before and after the calibration small.

In Oqtant, we can manipulate the evaporative cooling process by specifying the time evolution of power and frequency of an applied radio frequency field. The used parameters are shown in Table\,\ref{table_0}. To adjust the temperature, we simply changed the frequency at the last stage of the evaporative cooling process, from 0.02 MHz to 0.007 MHz. Table\,\ref{table_1} summarizes the prepared BEC parameters.

\begin{table}[h]
\caption{Evaporative process setting. The frequency value in the bracket shows the value after recalibration.
}
\begin{tabular}{c|ccccc}
time [ms] & 0 & 400 & 800 & 1200 & 1600 \\
power [mW] & 600 & 800 & 600 & 400 & 400 \\
frequency [MHz] & 21.12 & 12.12 & 5.12 & 0.62 & 0.02 (0.007)
\end{tabular}
\label{table_0}
\end{table}

\begin{table}[h]
\caption{Prepared BEC parameters.
The ``total'' column indicates the average and standard error (SE) over all days.
The ``before'' and ``after'' columns show, respectively, the average and SE before and after the recalibration.
The values in parentheses denote SEs. }
\begin{tabular}{c|ccc}
		     & total & before & after\\ \hline
Temperature [nK]   & 88(3) & 92(5) & 86(4) \\
Atom number & 18600(900) & 16500(1300) & 20000(1100)\\
Condensed Atom number & 8300(400) & 7000(500) & 9000(400)\\
\end{tabular}
\label{table_1}
\end{table}

\subsection*{Supplementary Note 2: Collective mode measurement}

After BEC preparation, we first apply $V_{\rm canc}$ to reduce unwanted dipole motion. The height of $V_{\rm canc}$ is 16.8 kHz. The barrier at the center of the trap was linearly loaded up to $V_0$ from 0 ms to 8 ms.

\begin{figure}[h]
\centering
\includegraphics[width=0.7\columnwidth]{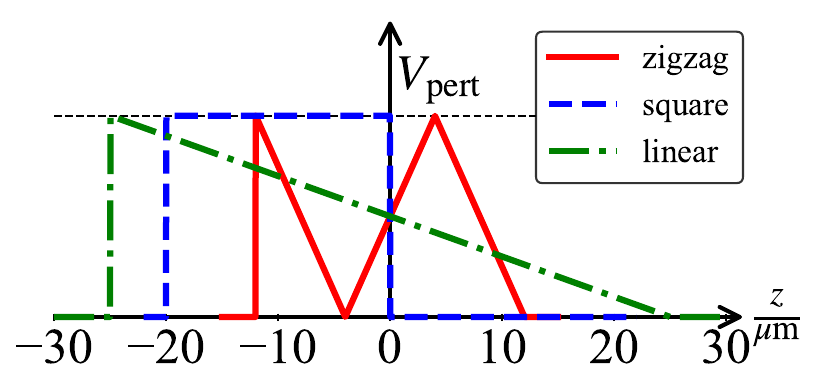}
\caption{Shapes of perturbation potentials to excite collective excitations. We call potentials depicted by solid, dashed, and dash-dotted lines ``zigzag'', ``square'', and ``linear'' perturbations, respectively.
}
\label{fig_type}
\end{figure}

As mentioned in the main text, we use three types of perturbation potentials to excite collective excitations, which are shown in Fig.\,\ref{fig_type}. We briefly note that the width of the ``zigzag'' potential is rather small compared to others, but is large enough to excite collective modes because the width of BECs, Thomas-Fermi radius, is approximately given by $10\mu$m.

Table\,\ref{table_2} summarizes which potential shape was used and which experiment is used for which frequency estimation. In the $V_0 = 0$ kHz case marked by $^*$ in Table\,\ref{table_2}, we do not apply $V_{\rm canc}$. The reason is as follows: In the absence of the barrier, the lowest-frequency collective mode is equivalent to a dipole motion. Since the prepared BEC is already in dipole motion, we can measure its frequency without perturbation if we do not apply $V_{\rm canc}$. Although a weak perturbation potential ($V_{\rm pert} = 0.243$kHz) is applied, it is sufficiently weak that it is considered to have no significant effects on the lowest mode.

\begin{table}[h]
\caption{Table of experimental barrier parameters and which experiments correspond to which frequency estimations. The potential shape is defined in Fig.\,\ref{fig_type}. The ``before/after'' column indicates whether the experiment date is before or after the recalibration. The barrier height before the recalibration corresponds to 0.7605 times the height after calibration. For values before the recalibration, we show values with this modification. The ``lowest'', ``second'', and ``third'' columns, respectively, indicate whether this experiment was used to estimate the frequency of the lowest-frequency, second-lowest-frequency, and third-lowest-frequency collective excitations. In the $V_0 = 0$ kHz case marked by $*$, we do not apply $V_{\rm canc}$.
}
\begin{tabular}{cccccccc}
$V_0$ [kHz] & Potential shape & $V_{\rm pert}$ [kHz] & before/after & lowest & second & third \\ \hline
0.0$^*$ & zigzag & 0.243 & before & $\checkmark$ & &  \\
0.0 & zigzag & 2.43 & before & & $\checkmark$ & $\checkmark$ \\
0.7605 & zigzag & 2.43 & before & & $\checkmark$ & $\checkmark$  \\
1.217 & zigzag & 2.43 & before & & $\checkmark$ & $\checkmark$ \\
1.673 & zigzag & 2.43 & before & & $\checkmark$ & $\checkmark$  \\ \hline
1.0 & linear & 1.68 & after & $\checkmark$ & &\\
1.5 & linear & 6.38 & after & & $\checkmark$ & $\checkmark$ \\
1.5 & square & 0.5 & after & $\checkmark$ & &\\
2.0 & square & 0.5 & after & $\checkmark$ & &\\
2.5 & linear & 3.6 & after & & $\checkmark$ & $\checkmark$ \\
2.5 & square & 0.5 & after & $\checkmark$ & & \\
3.0 & linear & 3.6 & after & & $\checkmark$ & $\checkmark$ \\
3.0 & square & 0.5 & after & $\checkmark$ & & \\
3.5 & square & 0.5 & after & $\checkmark$ & $\checkmark$ & $\checkmark$
\end{tabular}
\label{table_2}
\end{table}


\onecolumngrid

\end{document}